\documentclass[10pt,aps,prl]{revtex4}
\usepackage{graphicx,amssymb,color}
\usepackage{amsmath}
\usepackage{epstopdf}

\begin{document}
\title{One-dimensional sawtooth and zigzag lattices for ultracold atoms}

\author{Ting Zhang}
\author{Gyu-Boong Jo$^*$}

\affiliation{Department of Physics, \\Hong Kong University of Science and Technology, Clear Water Bay, Hong Kong, China.}

\email{gbjo@ust.hk}



\begin{abstract}
We describe tunable optical sawtooth and zigzag lattices for ultracold atoms. Making use of the superlattice generated by commensurate wavelengths of  light beams, tunable geometries including zigzag and sawtooth configurations can be realised. We provide an experimentally feasible method to fully control inter- ($t$) and intra- ($t'$) unit-cell  tunnelling in zigzag and sawtooth lattices. We analyse the conversion of the lattice geometry from zigzag to sawtooth, and show that a nearly flat band is attainable in the sawtooth configuration by means of tuning the lattice parameters. The bandwidth of the first excited band can be reduced up to 2$\%$ of the ground bandwidth for a wide range of lattice setting. A nearly flat band available in a tunable sawtooth lattice would offer a versatile platform for the study of interaction-driven quantum many-body states with ultracold atoms.
\end{abstract}
\maketitle





\flushbottom
\maketitle

\thispagestyle{empty}


Recent experimental and theoretical advancements in optical lattices hold out a promise of  engineering the band structures, the amplitude and the phase of inter-site tunnelling energies, and on-site interactions in the Hubbard regime~\cite{Bloch:2008gl,Morsch:2006em}. Furthermore, the highly tunable optical lattices \cite{Tarruell:2011wo,Luhmann:2014ud,Jo:2012br} would realise unconventional band structures that are not easy to study in solid-state materials, including  flat bands with completely quenched kinetic energies or topological bands. Flat band occurs in a variety of condensed-matter systems, ranging from the Landau levels of an electron gas\cite{Snoke:2008tj}, edge states of graphene~\cite{Nakada:1996us} to unconventional superconductors~\cite{Schnyder:2011kf}. Such flattened band structures significantly enhance interaction effects and show a wide range of interaction-driven  many-body phenomena from supersolidity \cite{Huber:2010bc} to flat-band ferromagnetism \cite{Tanaka:2007tb,Wu:2007iz}. Intriguingly, the divergence in the density of states at the energy of the flat band prevents natural ordering in the system\cite{Sun:2011dk,Wu:2007iz}, and the question still remains whether or not  Bose-Einstein condensation is stable in a flat band~\cite{You:2012cr,2015arXiv150505652B}. In addition, the interaction-induced localised state reveals crystalline phases in  fractional-filled bosons and fermions in a flat band \cite{Huber:2010bc} and more interestingly topological flat bands with nonzero Chern number open a new way for realising a fractional quantum hall state without Landau levels~\cite{Neupert:2011db,Sun:2011dk,Tang:2011by}.

In the ultracold atomic system, a bulk flat band appears though the whole system when the hopping matrix elements are destructively interfered such as in kagome \cite{Jo:2012br}, Lieb \cite{Noda:2014wk}, sawtooth\cite{Cai:2013tw,Huber:2010bc} lattices. However, it has not been experimentally explored in great depth so far how the flat dispersion affects the ground state of bosons and fermions except the very recent work in which the decay of the bosons from the flat excited band has been studied~\cite{Taie:2015tj}. This is because the flat band is usually not a ground band in a typical optical lattice with negative tunnelling energy~\cite{Jaksch:1998toa}. Indeed, the surface flat band has been also theoretically investigated in optical checkerboard lattices when the system is topologically nontrivial \cite{2015PhRvA..91c3604P}.  Intriguingly, a nearly flat band can be also obtained by including the spin degrees of freedom in the presence of the spin-orbit coupling~\cite{Zhang:2013fy,Lin:2011hn, 2014PhRvL.112k0404L} or the dipolar interactions~\cite{Yao:2012fn}. We note similar flattened quartic dispersion is also available in the shaken lattice \cite{Parker:2013ek}.

In this work, we describe a one-dimensional tunable optical zigzag lattice \cite{2013PhRvA..88e3625D,Greschner:2012va}  and then analyse the geometric conversion from a zigzag into a sawtooth optical lattice that provides a nearly flat band~\cite{Cai:2013tw,Huber:2010bc}. Especially, a sawtooth lattice has attracted significant interest due to the flat band structure but the creation of tunable sawtooth geometry has not been proposed to our knowledge. Our approach relies on a tunable superlattice generated by commensurate wavelengths that are red-detuned to the principle optical transition of atoms.  We first demonstrate a method to fully control tunnelling matrix elements in the zigzag configuration. The full control  of zigzag lattice is the critical step to access a nearly flat band in a sawtooth lattice  which is converted from the original zigzag configuration. Only for certain lattice configuration satisfying  the relation $t'/t = \sqrt{2}$ between inter-($t$) and intra-($t'$) unit-cell tunnelling energies, a flat band shows up in the sawtooth geometry. We analyse the band structure of the sawtooth and zigzag lattices for various lattice settings and provide a recipe for obtaining a nearly flat band in the sawtooth configuration. Besides the sawtooth lattice containing a flat band, a one-dimensional zigzag chain \cite{2013PhRvA..88e3625D,Greschner:2012va} also reveals particularly rich physics due to the interplay between frustration imposed by lattice geometry and two (three)-body interactions. Indeed, a chiral bosonic superfluid is expected in a zigzag lattice, which may become a Mott insulator even at  small interactions~\cite{Greschner:2012va,2013PhRvA..88e3625D}. 

\begin{figure}[tb]
\begin{center}
\includegraphics[width=4.5 in]{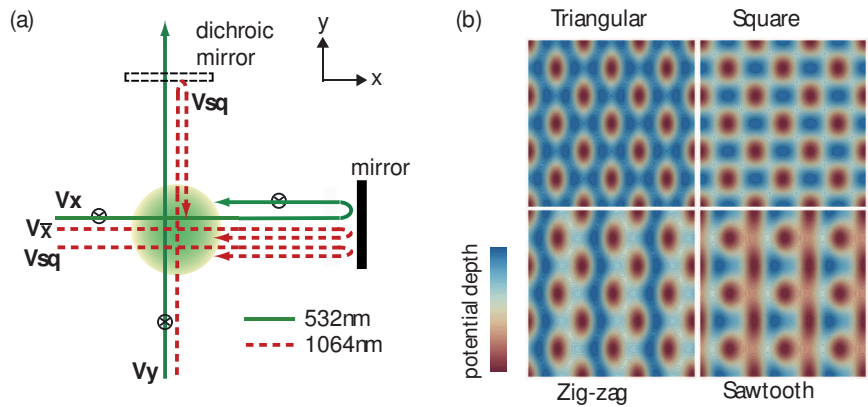}
\caption{Schematics of the optical supernalttice generating zigzag and sawtooth lattices. (a) Three bichromatic light beams with potential depth of $V_x$, $V_y$ and $V_{\overline{x}}$ form a tunable zigzag optical lattice with the two-dimensional potential shown in (b). Adding a two-dimensional square lattice with the depth of $V_{sq}$, the lattice is converted to the sawtooth geometry. (b)  Two-dimensional triangular and square lattices are also attainable without LW lattice ($V_{\overline{x}}$) when  $V_x=V_y$ and $16V_x=V_y$ respectively. A typical zigzag or sawtooth lattice is shown in (b) with the setting of ($V_x$,$V_y$,$V_{\overline{x}}$,$V_{sq}$)=$(1,1,1,0)$ or $(1,1,1,1) \times E_R$. In this case, the time phase is set to $\phi$= 0, $\phi_{sq,y}=0$, and $\phi_{sq,x}=-\pi$.
}
\label{fig:schematic}
\end{center}
\end{figure}

\paragraph{Creation of optical zigzag and sawtooth lattices} An array of one-dimensional zigzag lattices can be created in  two-dimensional superlattice geometry. Here,  the superlattice is generated by the incoherent superposition of bichromatic lattice beams. We assume that both short-wavelength (SW) and long-wavelength (LW) lattices are red-detuned from the principle optical transition of the atom and commensurate with each other. For example, the wavelength of 532nm (SW) and 1064nm (LW) lights are employed for ytterbium atoms with the principle optical transition of 399~nm~\cite{Fukuhara:2007wg,Takasu:2003we}. The SW lattice consists of the three intersecting beams of the same optical frequency along x- and y- directions (see Fig.~\ref{fig:schematic}). All lights are vertically polarised perpendicular to the lattice plane, which results in the SW lattice potential (say $V_{SW}$). Additionally, the LW lattice ($V_{\overline{x}}$) is added to decouple a y-directional zigzag chain from the adjacent one. We note other geometries such as a triangular or a square lattice can be also generated as described in  Fig.\ref{fig:schematic} (b), but we focus on the zigzag and sawtooth lattices for the rest of the discussion. Finally, additional square lattice generated by the 1064nm light  is incoherently added  with the maximum trap depth of $V_{sq}$ in order to convert the zigzag  to the sawtooth lattice.  Here, the frequency difference (e.g. at least 10~MHz or larger) between $V_{\overline{x}}$ and $V_{sq}$ laser fields is asuumed to minimise the interference effect.
 
With all beams in hand, the total optical potential is given as:
\begin{eqnarray}
V(x,y) 
&=&-[V_x \cos^2(kx)+\sqrt{V_x V_y} \cos{kx} \cos({ky+\phi})]  
-{V_{\overline{x}}}\cos^2({\frac{k}{2}x-\frac{\overline{\phi}_{x2}}{2}})-V_{sq}\sum_{i=x,y}[ \cos^2(\frac{k}{2}i+\frac{\phi_{sq,i}}{2})]
\label{pot}
\end{eqnarray}

where time-phase of the short-wave lattice is $\phi=\phi_{y1}-\phi_{x1}-\frac{\phi_{x2}}{2}$, $\phi_{i,1}$  the phase of the light at the position of atoms, $\phi_{i,2}$ the accumulated optical phase for the round trip from the atom where $i \in{x,y}$. Here, the lattice depth $V_{\alpha}$ where $\alpha=\{x,y,\overline{x},sq\}$  is the depth of the single-beam  normalised by the recoil energy for the short-wavelength $E_R=\frac{\hbar^2 k^2}{2m}$ where the wavenumber $k=\frac{2\pi}{\lambda}$ for $\lambda=532$nm. The time-phase of the lattice beam can be actively stabilised by means of controlling optical paths or the laser frequency. ~\cite{Jo:2012br,Tarruell:2011wo,Taie:2015tj}.  
For the rest of the discussion, we shall assume all optical phases are fixed as $\phi$= 0, $\overline{\phi}_{x2}$=$\pi/2$, $\phi_{sq,y}=0$, and $\phi_{sq,x}=-\pi$, but only each lattice depths ($V_x,V_y,V_{\overline{x}}$, $V_{sq}$) are varied.  





\begin{figure}[tb]
\begin{center}
\includegraphics[width=5.5 in]{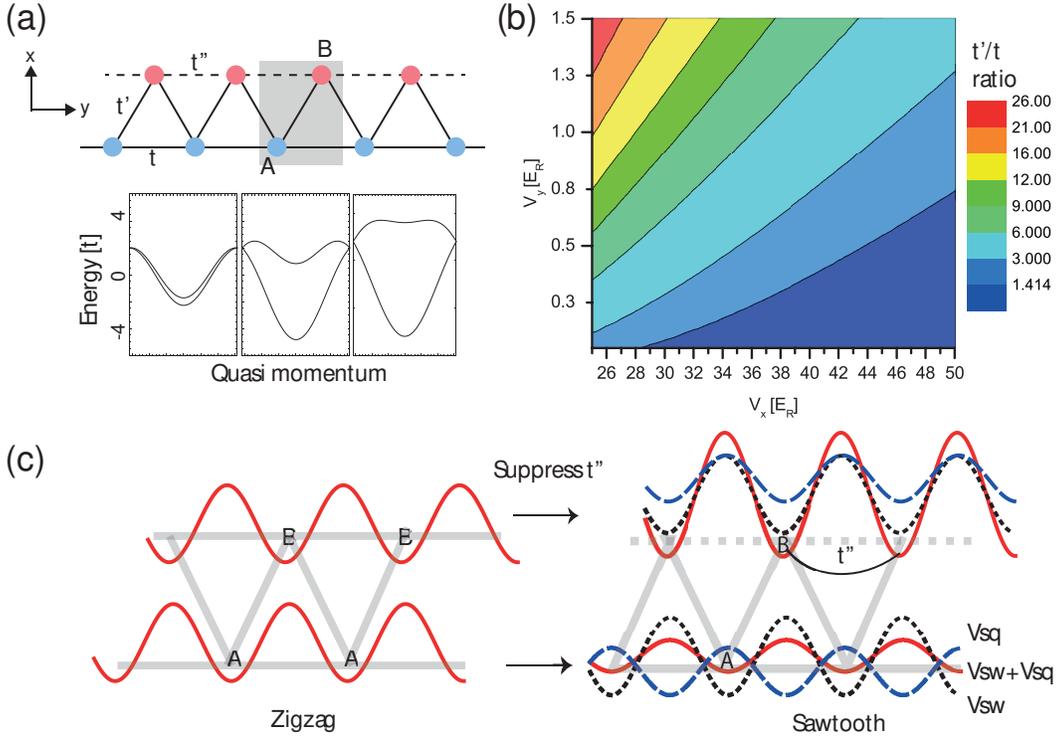}
\caption{Zigzag and sawtooth lattices. (a) A one-dimensional zigzag lattice consists of two sub-lattice sites $A$ and $B$ in a unit cell (gray in (a)) with tunnelling eneergies ($t$,$t'$ and $t''$). The dispersion of zigzag lattice in the case of $\epsilon_A=\epsilon_B$ and $t^{''}=t$ is shown for various cases: $t'<\sqrt{2}t$, $t'=\sqrt{2}t$, and $t'>\sqrt{2}t$ respectively (from left to right). (b) The ratio between inter-($t$)  and intra-($t'$) unit-cell tunnelling is calculated as a function of Vx and Vy in the zigzag lattice. (c) The zigzag lattice is converted to a sawtooth lattice when the tunnelling $t''$ is suppressed. The square lattice with the depth of $V_{sq}$ increases the barrier between the potential wells $B$ but decreases the one between the potential wells $A$. The lattice potential of ($V_{sw}+V_{sq}$), described by the solid line, exhibits sawtooth configuration.} \label{fig2}
\end{center}
\end{figure}


\paragraph{Zigzag and sawtooth lattices}
To achieve a sawtooth lattice with a nearly flat  band, we begin by considering  an alternative lattice structure which is easier to obtain, a zigzag lattice. In our optical potential  (eq. (\ref{pot})), if we set $V_{sq}=0$, then a zigzag lattice is at hand. The zigzag  lattice has the two lowest energy $s$-orbital bands corresponding to two basis lattice sites, labeled by $A$ and $B$, within the unit cell (see Fig. \ref{fig2}).  The inter-unit-cell tunnelling between neighbouring A (or B) wells is defined as  $t$ (or $t''$), and the intra-unit-cell tunnelling between A and B is $t'$.  We employ a large potential depth of the LW lattice $V_{\overline{x}}=25  E_R$ to isolate a one-dimensional array in vertical direction, and therefore the coupling between adjacent one-dimensional zigzag lattices is negligible.  With symmetric zigzag lattice configuration, we have $t''=t$. In this case, the band structure is determined by inter- ($t$, $t''$) and intra- ($t'$) unit-cell tunnelling energies as shown in Fig. \ref{fig2} (a) for variable tunnelling ratio $t'/t$. 

To search for a flat band in the sawtooth lattice, let us first consider the particular ratio of $t'/t$ required to support a flat band in a tight-binding model. A sawtooth lattice with a flat band will be obtained if we make $t''=0$ , and the relation $t'=\sqrt{2}t$ is satisfied~\cite{Huber:2010bc}.  These two conditions can be achieved by varying the parameters of the optical potential such as $(V_x,V_y,V_{\overline{x}},V_{sq})$.  To achieve the first condition of $t''=0$, we introduce the square lattice of the potential depth $V_{sq}$ as described below, and then we vary the ratio between   $V_x$ and $V_y$ to satisfy the second requirement of $t'/t= \sqrt{2}$ in the sawtooth lattice.

Here we first demonstrate the way to fully control the tunnelling ratio $t'/t$ in the zigzag case before we discuss the transform of the lattice into the sawtooth configuration. As it is difficult to obtain an obvious relation connecting the potential parameters $V_x$ and $V_y$ with the tunnelling energy $t$ and $t'$, we first calculate the Bloch state of the two-dimensional superlattice using the plane-wave expansion method with variable lattice settings~\cite{MPMarder:2000ta,Walters:2013tm} . Then, we perform least-square fitting of the band structure along the y-direction by the one-dimensional tight binding model of the zigzag lattice.  The  Hamiltonian for the zigzag lattice in the second quantised form is given:
\begin{eqnarray}
 H=-t'\sum_{i} ( \hat{a}^{\dag}_{i,A}\hat{a}_{i,B}+\hat{a}^{\dag}_{i,A}\hat{a}_{i-1,B})-t\sum_{i} ( \hat{a}^{\dag}_{i,A}\hat{a}_{i-1,A})-t''\sum_{i} ( \hat{a}^{\dag}_{i,B}\hat{a}_{i-1,B})+h.c
\end{eqnarray}
where $\hat{a}_{i,s}$ is the annihilation operator at the site $s \in \{A,B\}$ in the $i$-th unit cell.
Then in the momentum-space basis $\hat{\psi}_k=\{\hat{a}_{k,A},\hat{a}_{k,B}\}$, the Hamiltonian  is diagonalised giving the tight-binding matrix $H_{TB}$ of the zigzag lattice.
\begin{eqnarray}
H_{TB}=\sum_{k} \hat{\psi}_k^{\dag}\begin{pmatrix} \epsilon_A + 2t\cos(ka) & t'(1+e^{ka}) \\ t'(1+e^{-ka}) & \epsilon_B + 2t''\cos(ka) \end{pmatrix} \hat{\psi}_k^{T}
\label{tbeq}
\end{eqnarray}

where $\epsilon_{A,B}$ denotes the lowest eigen-energy of the quantum well $A$,$B$ and $a$ the lattice constant.

Since $\epsilon_A=\epsilon_B$ and $t''=t$ are naturally satisfied in the zigzag lattice we consider here, the tight binding dispersion for a zigzag lattice is then simplified to:
\begin{eqnarray}
 \epsilon(k)=2t\cos(ka) \pm t'\sqrt{2(1+\cos(ka))}
\label{eq4}
\end{eqnarray}


This dispersion is quite sensitive to the ratio $t'/t$ as shown in Fig.\ref{fig2} (a). Therefore fitting the band structure obtained by the plane-wave expansion method with the eq.(\ref{eq4}), we obtain the ratio of $t'/t$  from the potential parameters $V_x$ and $V_y$ as shown in Fig. \ref{fig2} (b).  To fulfil the condition  $t'/t=\sqrt{2}$, the ratio of $V_x/V_y$ should be extremely high around the order for about $\sim$100.  In this case,  the hopping integrals are quite sensitive to the overlap between Wannier functions.  Indeed, they are exponentially dependent on space coordinates, and therefore the hopping elements are very sensitive to the separation between the associated potential wells. For current form of the potential well, the separation between two neighbouring $A$ wells that is about  $\sqrt{2}$ times larger than the separation between neighbouring $A$ and $B$ wells, and therefore it will generate an extremely large $t'/t$   ratio.  To reduce this ratio down to $t'/t\sim \ensuremath{\operatorname{O}}(1)$, we have to modify the shape of the potential wells reducing  $t'$ and enhancing $t$ .  So it’s necessary to increase $V_x/V_y$ ratio, making the potential wells longer and narrower.  This method reduces the overlap between A and B wells, but enhances the overlap between neighbouring A wells explaining  the reason for such a large value of $V_x/V_y$. 

\begin{figure}
\begin{center}
\includegraphics[width=3.8in]{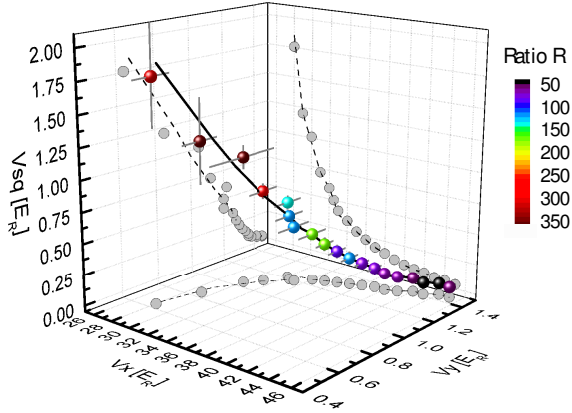}
\caption{Where a flattened band emerges in a sawtooth lattice. The ratio $R$ between the bandwidths for the ground and the first excited band is calculated for variable lattice depths $V_x$, $V_y$ and $V_{sq}$. In the diagram, the solid curve represents the lattice setting ($V_x,V_y,V_{sq}$) in which the flatness ratio $R$ is larger than 50.  The maximum flatness $R$ decrease as a function of the lattice depth $V_x$ as indicated by the colour scale. For each value of $V_x$, the bar around the date point indicates the characteristic range of ($V_y,V_{sq}$) in which the flatness $R$ is larger than 50. The solid and dashed lines are guides to the eye.}
\label{fig3}
\end{center}
\end{figure}

\paragraph{A nearly flat band in a sawtooth lattice}
After we tune the zigzag lattice with various values of $t'/t$ , the next step is to transform a zigzag lattice into a sawtooth lattice.  To suppress the tunnelling energy $t''$ appropriately, we now introduce a two-dimensional square lattice potential  with the depth of $V_{sq}$. The potential maximum (minimum) of the $V_{sq}$ lattice is aligned to the middle point of neighbouring $B$ ($A$) wells as described in Fig.~\ref{fig2} (c). Then, the square lattice with the depth $V_{sq}$ will generate an additional barrier between neighbouring B wells reducing  $t''$ significantly.  Because the barrier between neighbouring $A$ wells are reduced by $V_{sq}$, and therefore the $Vx/Vy$ should get  modified to take this effect into account.  From our analysis, making $t'/t\sim 10$ (in the zigzag lattice) is necessary to balance the effect of $V_{sq}$.  

\begin{figure}
\begin{center}
\includegraphics[width=5.5 in]{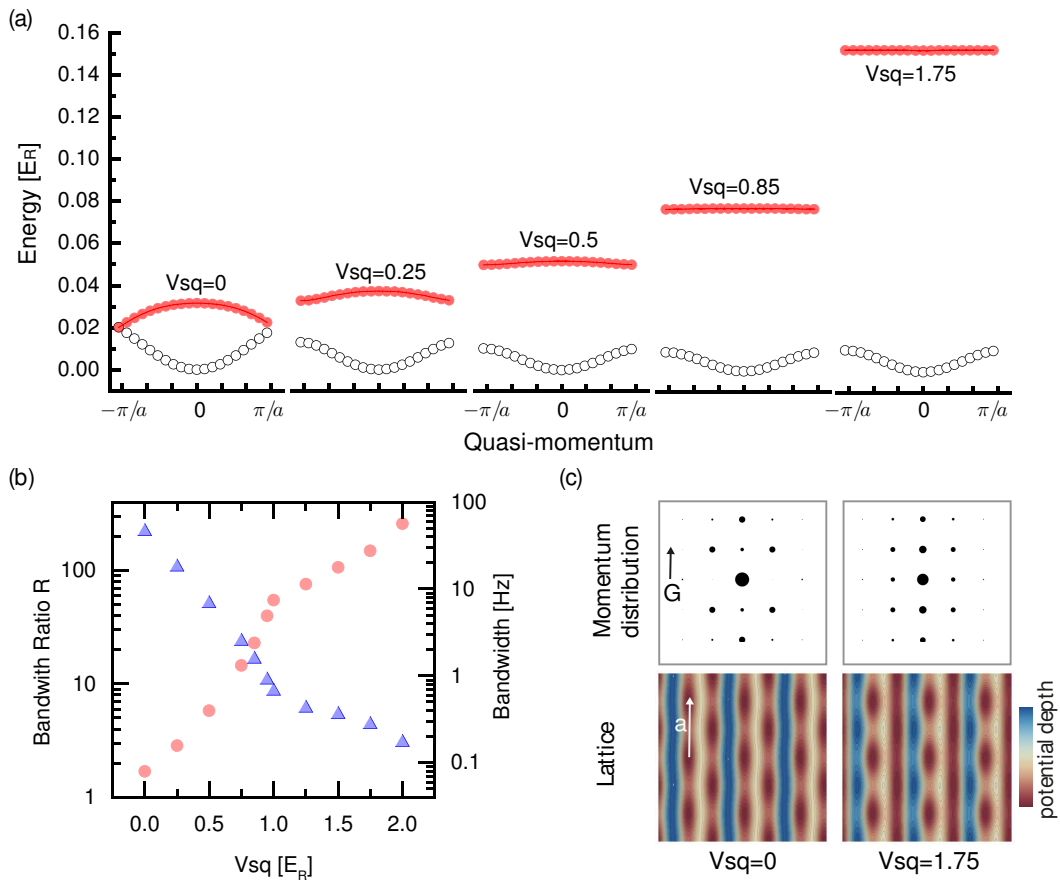}
\caption{Adiabatic transformation from the zigzag to the sawtooth lattice as $V_{sq}$  is increased. At the lattice setting of $(V_x,V_y,V_{\overline{x}},V_{sq})=(28,0.6,25,V_{sq})\times E_R$, the energies of the lowest two bands are calculated for $V_{sq}=\{0,0.25,0.5,0.85,1.75\}$$\times  E_R$ (left to right). (b) In this setting, the first excited band becomes flattened with bandwidths from $h\times$50~Hz to $h\times$0.2~Hz where the maximum flatness $R$ is $\sim$300. Here the bandwidth is calculated for ytterbium atoms as an example. (c) The real space optical potential and its corresponding momentum distribution are shown. The momentum distribution is calculated from the Fourier transform of the single-particle ground state in the real space. In the momentum distribution, the area of the black dot reflects the fractional population at the reciprocal site and the \textbf{G} and \textbf{a} denote the reciprocal and the lattice vectors of the one-dimensional zigzag lattice respectively.}
\label{fig4}
\end{center}
\end{figure}

Indeed we investigate the dependance of the flatness of the first excited band on the lattice setting $(V_x,V_y,V_{sq})$. To measure the flatness of the first excited band, we define the ratio between the bandwidths of the ground and the first excited band as $R$, and numerically calculate $R$ as a function of $Vx$, $Vy$, and $V_{sq}$ as shown in Fig.~\ref{fig3}. Since the ratio $V_x/V_y$ must be large to obtain $t'/t\sim \ensuremath{\operatorname{O}}(1)$ in the sawtooth lattice, we scan the potential depth $V_y$ around $V_y\sim E_R$ setting the potential depth $V_x$ from 28$E_R$ to 46$E_R$. As $V_x$ decreases, the ratio $t'/t$ increases (see Fig.~\ref{fig2} (b)). Therefore, the relatively deep square lattice potential of  $V_{sq}$ is required to keep the small value of  $t'/t$ for achieving a nearly flat band. From our numerical analysis, we find  the flatness of  R$>$50 can be easily achieved for a wide range of the lattice setting. In Fig.~\ref{fig3}, we provide 
a set of parameters $(V_x,V_y,V_{sq})$ where the flatness $R$ is at least over 50 from $V_x=28E_R$ and $V_x=46E_R$.

For a real flat band in a sawtooth lattice, however, it’s necessary to set the tunnelling $t''$ to be exactly zero.  This requires the barrier between neighbouring $B$ sites to be as large as possible, and therefore we should increase $V_{sq}$.  However the increase of $V_{sq}$ also has  consequence in reducing the barrier between neighbouring $A$ sites.  So with the presence of the square lattice,  $A$ and $B$ potential wells are no longer identical to each other.  This will give rise to slight difference in eigenvalues of the two potential wells $\epsilon_A$ and $\epsilon_B$ causing that the first excited band becomes dispersive (see eq.(\ref{tbeq})). Therefore the presence of the square lattice with $V_{sq}$ suppresses the tunnelling $t''$ but introduces the finite dispersion of the first excited band at the same time. Those limitations are prominent at the small value of $V_x \to 28 E_R$ in Fig.~\ref{fig3}, and thus the bandwidth ratio between the ground and the first excited bands has a maximum value of $R<$400. For a larger value of $V_x \to 46 E_R$, the first excited band become flattened at the small value of $V_{sq}$. In this case, the tunnelling $t''$ cannot be completely suppressed causing a small dispersion (or $R\sim 50$) of the band. This limitation can be readily solved by employing the lattice shaking technique which allows us to control the tunnelling $t$ and $t'$ independently~\cite{Lignier:2007du,Struck:2011is}. 


Finally we demonstrate how the dispersive band in the zigzag lattice is adiabatically changed into a nearly flat band in the sawtooth lattice. As an example in Fig.\ref{fig4} (a), the two lowest energy bands are shown as $V_{sq}$  increases with the lattice setting of $(V_x,V_y,V_{\overline{x}}, V_{sq})=(28,0.6,25, V_{sq})\times E_R$. The first excited band becomes flatter as the lattice geometry is changed from the zigzag to the sawtooth limit.  Indeed, the bandwidth of the first excited band becomes $<h\times$1~Hz in this sawtooth limit where $h$ is the Planck constant. Considering a typical energy scale of on-site interaction energy at the intermediate lattice depth, a nearly flat band realised in this setting should reveal the effect of flat band associated with interactions. In addition, geometry of the undergoing lattice structure can be easily confirmed in an experiment with a bosonic superfluid occupying the ground state of the lattice potential. In Fig.~\ref{fig4} (c), the expected momentum distribution of bosonic atoms is shown for zigzag and sawtooth lattices, which is obtained from the calculation of the single-particle ground state.

In optical lattices, a flat band often emerges in either the second or third lowest energy band for sawtooth (or Lieb) and kagome lattices respectively. In general, one may shake the optical lattice to flip the sign of the tunnelling energy by which atoms can be directly loaded in the flat band~\cite{Lignier:2007du,Struck:2011is}. For bosons, atoms can be coherently transferred at the particular quasi-momentum (say $k$=0 point) by manipulating the energy level of the sub-lattice site~\cite{Taie:2015tj}. The bosonic atoms in the flat band may decay into the lowest band as observed in the Lieb lattice~\cite{Taie:2015tj}, but the lifetime of atoms in the flat band can be lengthened by increasing the band gap~\cite{Taie:2015tj}. The effect of the flat band for fermions, however, should be observable with minimal decay into the lowest band when the Fermi energy is controlled near the energy of the flat band.


In this work, we provide a recipe to realise a one-dimensional sawtooth optical lattice with ultracold atoms. The highly tunable superlattice proposed here would offer a flexible experimental platform for the study of a flat band which is attainable in the sawtooth geometry. By simply adjusting lattice parameters and therefore controlling intra- and inter-unit-cell tunnelling , the first excited band becomes flat up to 2$\%$ of the ground bandwidth for a wide range of the lattice setting. Our superlattice  yields the zigzag and sawtooth lattices for alkali-earth-like bosons and fermions (e.g. $^{174,173}$Yb and $^{84,87}$Sr isotopes), in which various interaction-driven phenomena can be addressed in a flat band.




\section*{Acknowledgements}
The work was supported by the Hong Kong Research Grants Council (Project No. ECS26300014).

%

\end{document}